\documentclass{EJAM}

\usepackage{epstopdf}

\usepackage{upgreek}
\usepackage[numbers,sort&compress]{natbib}

\newcommand{\pfr}[2]{\ensuremath{\frac{\partial #1}{\partial #2}}}
\newcommand{\pfi}[2]{\ensuremath{{\partial #1}/{\partial #2}}} 
\newcommand{\mb}[1]{\mathbf{#1}}
\newcommand{\ep}{\varepsilon}
\newcommand{\gb}[1]{\boldsymbol{#1}}
\newcommand{\mc}[1]{\mathcal{#1}}
 
\newcommand{\tr}{\mathrm{tr}\,}
\newsavebox{\astrutbox}
\sbox{\astrutbox}{\rule[-5pt]{0pt}{20pt}}

\begin{document}

\jnlDoiYr{2026}
\doival{10.1017/xxxxx}

\lefttitle{Rajamanickam P.}

\papertitle{RESEARCH ARTICLE}

\title{Landau--de Gennes corrections to the Oseen--Frank limit: Anchoring-induced tilt modes}

\author{Prabakaran Rajamanickam}

\affil{Department of Mathematics \& Statistics, University of Strathclyde, Glasgow G1 1XQ, United Kingdom\\ Email: \href{mailto:prabakaran.rajamanickam@strath.ac.uk}{prabakaran.rajamanickam@strath.ac.uk}}


\begin{abstract}
An asymptotic analysis of the Landau--de Gennes framework is performed to compute higher-order 
corrections to the Oseen--Frank limit in a bounded three-dimensional domain under appropriately 
scaled surface anchoring energy. A systematic decomposition of the order parameter tensor into 
three mutually orthogonal subspaces—the uniaxial scalar, geometric tilt vector, and transverse 
anisotropy tensor—reveals that the leading $\mc O(\ep)$ correction to the Oseen--Frank director 
field $\mb n_0(\mb x)$ is dominated by a non-vanishing tilt field $\mb p_1(\mb x)$, where 
$\ep$ represents the ratio of the nematic coherence length to the characteristic domain 
size. This macroscopic variation constitutes a soft mode released from the boundary once the surface 
anchoring energy is retained at its physical scaling rather than driven to a formally infinite strength. 
We show that this tilt field is governed in the bulk by the linear homogeneous Jacobi equation, 
$\mc{J}_{\mb n_0}(\mb p_1) = \mb 0$, subject to a non-trivial, anchoring-driven Dirichlet boundary 
condition, where $\mc{J}_{\mb n_0}$ is the on-shell Jacobi operator of the harmonic map $\mb n_0$ 
on $\mathbb{S}^2$. The Oseen--Frank harmonic map and its on-shell Jacobi field are accompanied at 
$\mc O(\ep^2)$ by an off-shell correction to both the uniaxial scalar and the transverse anisotropy 
tensor, passively induced by the system's elastic non-uniformity $(\nabla\mb n_0 \neq \mb 0)$ coupled 
with the boundary-driven tilt $(\mb p_1 \neq \mb 0)$. Through $\mc O(\ep^2)$, the tilt enters the energy only through surface terms, not the bulk, providing a pathway for the system to lower its energy at the wall.  Under the conventional benchmark of 
rigid Dirichlet conditions, this response is annihilated outright, demonstrating that asymptotic corrections 
built upon infinite energy barriers obscure the underlying physics of anchoring-driven tilt modes.
\end{abstract}


\begin{msc}
{\it Primary} - 35B40;
{\it Secondary} - 35J57, 76A15, 49S05
\end{msc}

\maketitle

\section{Introduction}\label{sec:intro}

The macroscopic configurations and equilibrium topologies of nematic liquid crystals are traditionally described within the framework of the Oseen--Frank continuum theory, which delineates the spatial variations of a unit vector field $\mb n(\mb x)$ representing the local molecular director axis~\cite{de1995physics}. While this vector formulation enjoys widespread success when applied to 
large macroscopic regimes, its validity inevitably breaks down in regions characterised by sharp structural 
variations, most notably within the core neighbourhoods of singular defects and across highly localised 
boundary layers. The more comprehensive Landau--de Gennes framework circumvents these fundamental 
limitations by introducing a symmetric, traceless tensor order parameter $\mb Q$. This expanded 
formulation naturally incorporates fluctuations in the scalar degree of orientation, accounts for the 
phenomenon of localised core melting, and accommodates the emergence of local biaxiality.

It has been rigorously established that the Landau--de Gennes theory asymptotically approaches the classical Oseen--Frank continuum limit in regions remote from the singularities of $\mb n$~\cite{majumdar2010landau}, governed by the small dimensionless parameter $\ep = \xi / h \to 0$, wherein $h$ represents the 
characteristic linear dimension of the domain and $\xi$ denotes the intrinsic nematic coherence length. A frequent and mathematically convenient simplification invoked in the asymptotic analysis of this singular limit is the imposition of rigid Dirichlet boundary conditions at the bounding walls. Although such 
conditions are correct in a leading-order sense, higher-order profiles within the boundary layer are 
frequently obscured in the standard treatment. The enforcement of a strict Dirichlet condition effectively 
locks the director field rigidly at the interface across all asymptotic orders. Consequently, the leading-order 
corrections generated by the Landau--de Gennes framework are artificially confined to the bulk, manifesting 
purely as high-order eigenvalue distortions~\cite{di2020landau,wang2026convergence,nguyen2013refined}, specifically, 
as localised scalar melting and sub-principal biaxial corrections surfacing at $\mc O(\ep^2)$. In a recent 
investigation~\cite{rajamanickam2026strong}, we demonstrated that a substantially larger correction to the 
director field itself is realised within a reduced Landau--de Gennes framework when the surface anchoring 
energy is maintained at its physically appropriate scaling. The reduced framework, which has received a lot of attention recently~\cite{luo2012multistability,majumdar2016multistable,han2020reduced,rajamanickam2026nematic}, is useful to model nematic equilibria in confined systems such as planar bistable nematic device~\cite{tsakonas2007multistable} and the zenithally bistable nematic device~\cite{spencer2010zenithal}.

In the present work, we extend this asymptotic treatment to the full, physically comprehensive three-dimensional 
framework. Taking the classical Rapini--Papoular surface anchoring energy as a primary illustration---retained 
here at its appropriate physical scaling rather than driven to the formally infinite strength implicit in the 
Dirichlet assumption---we demonstrate that the anchoring-driven relaxation of the boundary transforms the 
structural response of the system entirely. Rather than being restricted to the passive, higher-order bulk 
eigenvalue variations recovered under rigid boundary constraints, the  freedom of the boundary profile 
acts as a direct source term that liberates macroscopic, geometric \textit{Goldstone-like tilt modes} in the director 
field. Because the boundary responds  to the prevailing elastic stresses at the wall, these tilt 
perturbations emerge at a lower asymptotic threshold than any previously identified correction, manifesting 
as the dominant $\mc O(\ep)$ modification to the continuum field.

\section{The Landau--de Gennes framework}\label{sec:framework}

Consider a bounded three-dimensional domain $\Omega$, filled with a nematic liquid crystal sample, enclosed by a piecewise smooth boundary $\partial\Omega$ of characteristic linear dimension $h$. The molecular orientation at the bounding surface is governed by an anchoring condition which, for mathematical convenience and clarity of exposition, we model by means of a Rapini--Papoular surface energy functional appropriate for non-degenerate directional alignment~\cite{mottram2014introduction}. The macrostate of the medium is characterized by the tensor order parameter $\mb Q$, which resides within the vector space $\mc S$ of symmetric, traceless $3 \times 3$ matrices:
\begin{equation}
    \mc S = \left\{ \mb Q \in \mathbb{R}^{3 \times 3} \mid \mb Q^T = \mb Q,\ \tr\mb Q=0 \right\}.
\end{equation}
Within the framework of the Landau--de Gennes theory, and invoking the standard one-elastic-constant approximation, the total free energy of the system is formulated as
\begin{align}
    F[\mb Q] = \int_\Omega \left[\frac{L}{2}|\nabla\mb Q|^2  +\frac{A}{2}\tr\mb Q^2 -  \frac{B}{3}\tr\mb Q^3 + \frac{C}{4}(\tr\mb Q^2)^2\right]dV 
   + \frac{W}{2}\int_{\partial\Omega}|\mb Q-\mb Q_b|^2 d\Sigma,\label{Feq0}
\end{align}
wherein $L$ and $C$ denote positive material constants, $B$ represents a positive cubic scaling coefficient, and $W$ is the surface anchoring strength. The thermotropic coefficient $A=A(T)$ varies linearly with the absolute temperature $T$. The tensor field $\mb Q_b$ prescribes the preferred, or easy, orientation of the order parameter at the boundary interface. The isotropic state corresponds identically to the vanishing configuration $\mb Q = \mb 0$, whereas the emergence of ordered nematic phases is characterized by $\mb Q \neq \mb 0$. In the ideal benchmark configuration of a spatially uniform medium remote from boundary constraints, the elastic and anchoring contributions in~\eqref{Feq0} vanish identically, whereupon the bulk free energy density is globally minimized by the homogeneous state
\begin{equation}
    \mb Q_{\text{homo}} = \begin{cases}
        \mb 0  & \quad \text{for  } A > B^2/27C,\\
        s_+ \left(\mb n_+\otimes\mb n_+-\tfrac{1}{3}\mb I\right)  & \quad \text{for  } A < B^2/27C,
    \end{cases}     
\end{equation}
where the bulk scalar order parameter is given by $s_+=(B+\sqrt{B^2-24AC})/4C$, and $\mb n_+\in \mathbb S^2$ represents an arbitrary unit director vector.

The spatial variations within the medium are naturally scaled by the intrinsic nematic coherence length $\xi(T)$ and the de Gennes--Kleman extrapolation length $l_{\text{ex}}$, defined respectively as\footnote{A more general formulation defines the nematic coherence length via the response function $\chi$ of the uniform, uniaxial medium such that $\xi \sim \sqrt{L\chi}$. Here, $\chi^{-1}=\left.\frac{d^2f_{\text{bulk}}}{ds^2}\right|_{s=s_+}=\frac{2}{3}A-\frac{4}{9}Bs_++\frac{4}{3}Cs_+^2$, which remains strictly positive for $A < B^2/24C$. In the deep nematic limit $A \to -\infty$, the scalar order approaches $s_+\to \sqrt{\tfrac{3}{2}|A|/C}$, yielding the asymptotic limit $\chi^{-1}\to \frac{4}{3}|A|$. For analytical brevity, we adopt the simpler definition $\xi^2=L/s_+^2 C$.}
\begin{equation}
    \xi(T)=\sqrt{\frac{L}{s_+^2C}}, \qquad  l_{ex}=\frac{L}{W}.
\end{equation}
Following the framework established in~\cite{rajamanickam2026strong}, we define the regime of strong surface anchoring by requiring that the ratio of these characteristic lengths satisfies the scaling condition
\begin{equation}
     \gamma = \frac{\xi}{l_{\text{ex}}} \sim \mc O(1).
\end{equation}
The pure Dirichlet boundary condition corresponds to the limit $\gamma\to \infty$. However, as argued in~\cite{rajamanickam2026strong}, $\gamma$ seldom becomes very large since $l_{ex}\sim 10^{-8}-10^{-5}$m~\cite{ravnik2009landau} and the coherence length is $\xi \sim 10^{-8}$m~\cite{de1995physics} even very close to the nematic–isotropic transition. Thus, the scaling $\gamma\sim \mc O(1)$ corresponds to the strongest anchoring condition achievable in practice.

To cast the system into a non-dimensional form, we rescale all spatial coordinates by the macro-dimension $h$, normalise the order parameters $\mb Q$ and $\mb Q_b$ by the bulk scalar value $s_+$, and scale the total energy functional by $s_+^2 Lh$. Under this transformation, the dimensionless free energy functional reduces to
\begin{align}
      \ep^2F[\mb Q] = \int_\Omega \left[\frac{\ep^2}{2}|\nabla \mb Q|^2 +\frac{\mc A}{2}\tr\mb Q^2 - \frac{\mc B}{3}\tr\mb Q^3+ \frac{1}{4}(\tr\mb Q^2)^2\right] dV + \frac{\ep\gamma}{2}\int_{\partial\Omega}|\mb Q-\mb Q_b|^2 d\Sigma, \label{freeenergy}
\end{align}
where the system metrics are defined by the dimensionless parameter groups
\begin{equation}
   \ep = \frac{\xi}{h}, \qquad \mc A = \frac{A}{Cs_+^2}\in\left(-\tfrac{2}{3},\tfrac{2}{3}\right), \qquad \mc B = \frac{B}{Cs_+}=3\mc A+2\in (0,4).
\end{equation}
Since $\mc A$ and $\mc B$ are related, they are used interchangeably; the entire physical range of $A\in(-\infty,B^2/24C)$ is now mapped to the interval $\mc A\in(-\tfrac{2}{3},\tfrac{2}{3})$. The corresponding Euler--Lagrange equations, derived by minimisation subject to the mathematical requirements of symmetry and tracelessness, take the form
\begin{equation}
    \ep^2\nabla^2 \mb Q = \gb\Lambda(\mb Q), \label{Qeq}
\end{equation}
where the non-linear bulk forcing term $\gb\Lambda(\mb Q)$ is defined by
\begin{equation}
    \gb\Lambda(\mb Q) = (\mc A + \tr\mb Q^2) \mb Q - (3\mc A+2) (\mb Q^2 - \tfrac{1}{3}\mb I\tr\mb Q^2).
\end{equation}
The boundary condition is given by
\begin{equation}
     \mb Q = \mb Q_b  -\frac{\ep}{\gamma}\frac{\partial \mb Q}{\partial\nu} \qquad \text{on} \qquad \partial\Omega, \label{BCs}
\end{equation}
where $\gb\nu$ denotes the outward unit normal vector to the boundary surface. The problem is thus seen to depend fundamentally upon three dimensionless parameters: the geometric scale ratio $\ep$, the reduced temperature $\mc A$, and the relative anchoring strength $\gamma$. In what follows, the preferred configuration at the boundary is assumed to be of purely uniaxial character, corresponding to a perfectly ordered homogeneous nematic interface,
\begin{equation}
    \mb Q_b = \mb n_b \otimes \mb n_b-\tfrac{1}{3}\mb I,
\end{equation}
where the surface director field $\mb n_b$ is explicitly prescribed. The asymptotic structure corresponding to the small-domain limit ($\ep \to \infty$) is detailed in Appendix~\ref{sec:small}.

\subsection{A representation for the $\mb Q$-tensor} \label{sec:tensor}

For the development of the subsequent asymptotic theory, it is advantageous to introduce a specific representation for the macrostate tensor $\mb Q$. This formulation systematically decomposes the $\mb Q$-tensor into three mutually orthogonal components defined with respect to a prescribed unit vector field $\mb n(\mb x)$. Manifestly, this geometric construction remains valid only within domains where the orientation field $\mb n$ is uniquely defined, and it undergoes a structural breakdown in regions containing topological singularities. Adopting this basis, we partition the order parameter tensor as
\begin{equation}
    \mb Q = \mb Q_{\parallel} + \mb Q_{\text{tilt}} + \mb Q_{\text{aniso}},
\end{equation}
wherein the constituent tensor fields are explicitly defined by
\begin{align}
    &\mb Q_{\parallel} = s \left(\mb n \otimes \mb n-\tfrac{1}{3}\mb I\right), \qquad  s = \tfrac{3}{2}\mb n^T\mb Q \mb n,\\
   &\mb Q_{\text{tilt}} = \mb p \otimes \mb n +\mb n\otimes \mb p, \qquad  \mb p = \mb P_\perp (\mb Q\mb n),\\
   &\mb Q_{\text{aniso}} = \mb P_\perp \mb Q\mb P_\perp-\tfrac{1}{2}\tr(\mb P_\perp\mb Q)\mb P_\perp , \qquad \mb P_\perp = \mb I-\mb n\otimes\mb n.
\end{align}
The five independent degrees of freedom inherent to the symmetric, traceless $\mb Q$-field are thus elegantly encoded through the \textit{uniaxial scalar} $s$, the \textit{tilt vector} $\mb p$, and the \textit{transverse anisotropy tensor} $\mb Q_{\text{aniso}}$. The last of these components, which is itself symmetric and traceless, is constrained to live entirely upon the local two-dimensional tangent plane, satisfying the contractive property $\mb Q_{\text{aniso}}\mb n=\mb 0$. Such a configuration space can invariably be parameterised as
\begin{equation}
    \mb Q_{\text{aniso}} = r\left(\mb m \otimes \mb m-\tfrac{1}{2}\mb P_\perp\right),
\end{equation}
which introduces a secondary scalar amplitude $r$ and a unit vector $\mb m$ strictly satisfying the transverse condition $\mb m\cdot\mb n=0$. By definition, the system conforms to the following set of mutual orthogonality relations:
\begin{align}
    \mb p\cdot\mb n=0, \quad\mb m\cdot\mb n=0, \quad \tr(\mb Q_{\parallel}\mb Q_{\text{tilt}})=0,  \quad \tr(\mb Q_{\text{tilt}}\mb Q_{\text{aniso}})=0, \quad \tr(\mb Q_{\parallel}\mb Q_{\text{aniso}})=0.
\end{align}
Conversely, the inner product of the two transverse vectors is generally non-vanishing, $\mb p \cdot\mb m \neq 0$, the relative orientation between $\mb p$ and $\mb m$ being dictated by the local mechanics and energetic requirements of the system under consideration. Although both fields are confined to the same tangent plane, a fundamental mathematical distinction exists in their tensor projections, namely
\begin{align}
    \mb Q_{\text{tilt}}\mb n = \mb 0 \qquad \text{if and only if}\qquad \mb Q_{\text{tilt}}=\mb 0,\quad 
    \text{whereas} \qquad \mb Q_{\text{aniso}}\mb n=\mb 0 \quad \text{identically}.
\end{align}
Furthermore, the physical description remains invariant under the two independent inversion choices $\mb n \sim -\mb n$ (coupled with $\mb p\sim-\mb p$) and $\mb m\sim -\mb m$, a property that faithfully mirrors the headless, non-polar symmetry of the liquid crystal molecules. To illustrate this representation concretely, let us align the reference director along the vertical axis, $\mb n(\mb x)=\mb e_z$. In the standard Cartesian basis $(\mb e_x,\mb e_y,\mb e_z)$, the components of the $\mb Q$-tensor expand into the matrix sum
\begin{align}
    \mb Q = \begin{pmatrix}
        -\tfrac{1}{3} s & 0  & 0 \\ 0 & -\tfrac{1}{3}s & 0 \\ 0 & 0 & \tfrac{2}{3}s
    \end{pmatrix} + \begin{pmatrix}
        0 & 0  & s\sin\theta\cos\phi \\ 0 & 0 & s\sin\theta\sin\phi \\ s\sin\theta\cos\phi & s\sin\theta\sin\phi & 0
    \end{pmatrix}  + \begin{pmatrix}
        \tfrac{1}{2}r\cos(2\psi) & \tfrac{1}{2}r\sin(2\psi) & 0 \\ \tfrac{1}{2}r\sin(2\psi) & -\tfrac{1}{2}r\cos(2\psi) & 0 \\ 0 & 0 & 0
    \end{pmatrix} ,
\end{align}
where the geometric tilt vector is identified as $\mb p=(s\sin\theta\cos\phi,s\sin\theta\sin\phi,0)^T$, and the transverse anisotropy axis is specified by $\mb m=(\cos\psi,\sin\psi,0)^T$.

Consider a localised sub-domain wherein the director field $\mb n$ maintains a uniform orientation. The physical interpretation of the tripartite representation follows naturally: (i) the scalar $s$ quantifies the degree of molecular alignment along the director axis $\mb n$, (ii) the vector $\mb p$ measures the physical tilt or angular deviation of the local director field with respect to $\mb n$, and (iii) the tensor $\mb Q_{\text{aniso}}$ isolates the residual degree of anisotropy within the transverse plane. While both $\mb p$ and $\mb m$ are embedded within the local tangent manifold of $\mb n(\mb x)$, they spring from distinct physical origins. The tilt vector is fundamentally geometric, intimately coupled to the macroscopic reorientation of the director axis; the transverse anisotropy, by contrast, is a material property of the order tensor, measuring the departure from uniaxiality, and it is also excited by geometric frustration, since strongly frustrated configurations tend to be biaxial. For example, in a spatially uniform uniaxial nematic phase, an unconstrained tilt vector is associated with the arbitrary rotational freedom of the director field on the unit sphere $\mathbb S^2$, thereby manifesting as a gapless Goldstone mode. Conversely, a non-vanishing transverse anisotropy tensor depends explicitly upon short-range molecular interactions and structural constraints. It must be noted that both $\mb Q_{\text{tilt}}$ and $\mb Q_{\text{aniso}}$ can generate net biaxiality within the total $\mb Q$-tensor. A pure tilt, isolated from the other components, is biaxial at second order in $|\mb p|$. A rigid rotation of the uniaxial state, which is uniaxial for all rotation angles, is recovered only when the tilt is accompanied by matching $\mathcal O(|\mb p|^2)$ corrections to the scalar $s$ and the anisotropy $r$. To leading order in the tilt, however, these corrections are negligible and a small tilt is indistinguishable from a rotation.

\section{The Oseen--Frank limit ($\ep\to 0$)}

The limit of vanishingly small coherence length, $\ep \to 0$, represents the regime of primary physical relevance, wherein the tensor order parameter $\mb Q$ asymptotically approaches a purely uniaxial state. The mathematical landscape of the Landau--de Gennes framework is known to exhibit significant complexity in this limit, characterised by a multiplicity of solution branches and critical points connected via unstable saddle configurations~\cite{han2020reduced,han2023multistability,han2021solution,yin2020construction}. The objective of the present analysis is not to chart these aspects of solution multiplicity or global thermodynamic stability, but rather to construct higher-order asymptotic corrections to the uniaxial state, given that a specific energy-minimising configuration has been selected as the baseline field.

To this end, we introduce the formal asymptotic expansion
\begin{equation} \label{Qexpansion}
    \mb Q = \mb Q_0 + \ep\mb Q_1 + \ep^2\mb Q_2 + \cdots,
\end{equation}
where the leading-order macrostate is delineated by the uniaxial profile $\mb Q_0=s_0(\mb n_0\otimes \mb n_0-\tfrac{1}{3}\mb I)$. In the bulk domain, the zeroth-order problem requires compliance with the homogeneous equilibrium relation
\begin{equation}
    \gb\Lambda(\mb Q_0)=\mb 0, \qquad \implies \quad s_0=1, \,\,  \mb Q_0=\mb n_0\otimes \mb n_0-\tfrac{1}{3}\mb I \qquad \text{in} \qquad \Omega.
\end{equation}
The Oseen--Frank limit thus manifests as a singular perturbation problem where the scalar order parameter $s_0$ is driven to unity throughout the bulk, save for localised regions containing topological defects where $s_0$ must necessarily vanish to preserve regular field configurations. In the lexicon of matched asymptotic expansions, the bulk 
regime wherein $s_0=1$ constitutes the \textit{outer region}. The present investigation is confined strictly to the analysis of corrections within this outer domain, under the assumption that the highly localised core structures of any interior defects are passively enslaved to these primary outer perturbations.

The bulk potential $f_{bulk}(\mb Q)=\tfrac{1}{2}\mc A \tr \mb Q^2 -\tfrac{1}{3}\mc B\tr \mb Q^3 +\tfrac{1}{4}(\tr\mb Q^2)^2$ can be written alternatively as
\begin{equation}
    f_{bulk}(\mb Q_0+\delta\mb Q) = f_{bulk}(\mb Q_0) + \tfrac{1}{2}\delta \mb Q:\gb\Lambda_{\mb Q_0}'(\delta\mb Q) + \tfrac{1}{6}\delta \mb Q:\gb \Lambda_{\mb Q_0}''(\delta\mb Q,\delta \mb Q) + \tfrac{1}{24}\delta \mb Q:\gb \Lambda_{\mb Q_0}'''(\delta\mb Q,\delta \mb Q,\delta \mb Q)
\end{equation}
where
\begin{align}
    &\gb\Lambda_{\mb Q_0}'(\mb X)=\left(\mc A+\tfrac{2}{3}\right)\mb X -\mc B\left(\mb Q_0\mb X+\mb X\mb Q_0-\tfrac{2}{3}\mb I\tr(\mb Q_0\mb X)\right) +  2\mb Q_0\tr(\mb Q_0\mb X),\\
     &\gb\Lambda_{\mb Q_0}''(\mb X,\mb Y)=-\mc B\left(\mb X\mb Y+\mb Y\mb X-\tfrac{2}{3}\mb I\tr(\mb X\mb Y)\right) +  2\tr(\mb X\mb Y)\mb Q_0 + 2\tr(\mb Q_0\mb X)\mb Y + 2\tr(\mb Q_0\mb Y)\mb X,\\
      &\gb\Lambda_{\mb Q_0}'''(\mb X,\mb Y,\mb Z)=  2\tr(\mb X\mb Y)\mb Z + 2\tr(\mb Z\mb X)\mb Y + 2\tr(\mb Z\mb Y)\mb X.
\end{align}
Substituting the asymptotic expansion~\eqref{Qexpansion}, i.e., $\delta\mb Q= \ep \mb Q_1+\ep^2\mb Q_2+\cdots$ yields the corresponding expansion for the bulk potential.

\subsection{The natural basis}
 The leading-order director field $\mb n_0(\mb x)$ provides a natural geometric framework to establish a triad of orthogonal basis tensors, $(\mb E_\parallel,\mb E_{\text{tilt}},\mb E_{\text{aniso}})$, which partition the total state space according to
\begin{equation}
    \mb Q = s \mb E_\parallel + \mb E_{\text{tilt}} + r \mb E_{\text{aniso}}.
\end{equation}
Here, the base components are defined explicitly by $\mb E_\parallel = \mb Q_0$, $\mb E_{\text{tilt}}=\mb p\otimes\mb n_0+\mb n_0\otimes\mb p$, and $\mb E_{\text{aniso}}= \mb m\otimes\mb m-\tfrac{1}{2}\mb P_\perp$, where $\mb P_\perp=\tfrac{2}{3}\mb I - \mb Q_0$ projects onto the local tangent manifold, and $\mb m$ is a transverse unit vector satisfying $|\mb m|^2=1$. The expansion of the global order parameter tensor~\eqref{Qexpansion} then maps directly onto the component expansions
\begin{align}
    s=1+\ep s_1+\cdots, &\qquad \mb p = \mb 0 + \ep\mb p_1 + \cdots, \\
    r=0+\ep r_1+\cdots, &\qquad \mb m = \mb m_0 + \ep\mb m_1 + \cdots
\end{align}
Consequently, the structural basis perturbations expand as
\begin{align}
    &\mb E_{\text{tilt}} = \ep\mb E_{\text{tilt}}^{(1)} +  \ep^2 \mb E_{\text{tilt}}^{(2)} + \cdots,\\
    &r\mb E_{\text{aniso}} = \ep r_1 \mb E_{\text{aniso}}^{(0)} + \ep^2\left[r_2 \mb E_{\text{aniso}}^{(0)} + r_1(\mb m_0\otimes\mb m_1+\mb m_1\otimes \mb m_0)\right] + \cdots,
\end{align}
where we introduce the notation $\mb E_{\text{tilt}}^{(j)}=\mb p_j\otimes\mb n_0+\mb n_0\otimes\mb p_j$ and $\mb E_{\text{aniso}}^{(j)}= \mb m_j\otimes\mb m_j-\tfrac{1}{2}\mb P_\perp$. As demonstrated in the subsequent analysis, the first-order transverse anisotropy vanishes identically, $r_1=0$, thereby establishing $\mb E_{\text{aniso}}^{(0)}$ with $|\mb m_0|^2=1$ as a robust invariant representation of the transverse sub-space up to order $\mc O(\ep^2)$.

\subsection{First order}
At the first asymptotic order, the governing  equations simplify to the linearised system
\begin{equation}\label{firstordereq}
   \gb\Lambda_{\mb Q_0}'(\mb Q_1)=\mb 0 \qquad \text{in} \qquad \Omega,
\end{equation}
The eigenvalues of the operator $\gb\Lambda'(\mb X)$ are readily evaluated by projection onto the decoupled orthogonal eigenspaces $(\mb E_\parallel,\mb E_{\text{tilt}}^{(1)},\mb E_{\text{aniso}}^{(0)})$, yielding the distinct spectral metrics
\begin{equation}
    \lambda_\parallel = \tfrac{2}{3}-\mc A, \qquad \lambda_{\text{tilt}}=0, \qquad \lambda_{\text{aniso}}=\mc B.
\end{equation}
Accordingly, by expressing the first-order perturbation in the natural basis, $\mb Q_1 = s_1 \mb E_\parallel + \mb E_{\text{tilt}}^{(1)} + r_1 \mb E_{\text{aniso}}^{(0)}$, the system~\eqref{firstordereq} reduces to the uncoupled algebraic relation
\begin{equation}
    s_1 \left(\tfrac{2}{3}-\mc A\right) \mb E_\parallel + 0\cdot \mb  E_{\text{tilt}}^{(1)} + r_1 \mc B\mb E_{\text{aniso}}^{(0)}=\mb 0.
\end{equation}
Given that the material coefficients satisfy $\mc A < \tfrac{2}{3}$ and $\mc B > 0$ within the nematic regime, we deduce that $s_1 = 0$ and $r_1 = 0$. This yields the foundational result that
\begin{equation}
     \mb Q_1= \mb p_1 \otimes \mb n_0 +\mb n_0\otimes \mb p_1,
\end{equation}
confirming that the first-order correction $\mb Q_1$ constitutes a pure, unconstrained tilt mode.

\subsection{Second order}
At the secondary asymptotic stage, the macroscopic elastic distortions interact with the lower-order fields, yielding the non-homogeneous equation system
\begin{equation}\label{secondordereq}
   \gb\Lambda_{\mb Q_0}'(\mb Q_2) = - \tfrac{1}{2}\gb\Lambda_{\mb Q_0}''(\mb Q_1,\mb Q_1)+ \nabla^2\mb Q_0, 
\end{equation}
wherein the structural forcing terms evaluate to
\begin{align}
    -\tfrac{1}{2}\gb\Lambda_{\mb Q_0}''(\mb Q_1,\mb Q_1) &= \left(\tfrac{3}{2}\mc A-1\right)|\mb p_1|^2 \mb Q_0 + \mc B\left(\mb p_1\otimes\mb p_1-\tfrac{1}{2}|\mb p_1|^2\mb P_\perp\right), \label{doublederi} \\
    \nabla^2\mb Q_0 &= (\nabla^2\mb n_0)\otimes\mb n_0 + \mb n_0\otimes(\nabla^2\mb n_0) + 2(\nabla\mb n_0)(\nabla\mb n_0)^T. \label{lap1}
\end{align}
Expanding the left-hand side of~\eqref{secondordereq} in terms of the baseline natural coordinates, $\mb Q_2 = s_2 \mb E_\parallel + \mb E_{\text{tilt}}^{(2)} + r_2 \mb E_{\text{aniso}}^{(0)}$, we obtain
\begin{equation}
     \gb\Lambda_{\mb Q_0}'(\mb Q_2) =  s_2 \left(\tfrac{2}{3}-\mc A\right) \mb E_\parallel+ r_2 \mc B\mb E_{\text{aniso}}^{(0)}.
\end{equation}
The linear operator does not generate a response within the geometric tilt subspace at this order, leaving the tilt component $\mb E_{\text{tilt}}^{(2)}$ uncoupled on the left-hand side.

To determine the structural corrections, we project the right-hand side forcing terms onto the same tripartite basis $(\mb E_\parallel,\mb E_{\text{tilt}}^{(2)},\mb E_{\text{aniso}}^{(0)})$. The contribution from the quadratic potential derivative is straightforward; from~\eqref{doublederi}, we isolate a component aligned with $\mb E_\parallel$, while the remainder maps onto the transverse anisotropy space owing to its property of vanishing under right-contraction with the director field $\mb n_0$. This decomposition is denoted as
\begin{equation}
     -\tfrac{1}{2}\gb\Lambda_{\mb Q_0}''(\mb Q_1,\mb Q_1) = \left(\tfrac{3}{2}\mc A-1\right)|\mb p_1|^2 \mb E_\parallel + \mc B\left\{\mb p_1\otimes\mb p_1-\tfrac{1}{2}|\mb p_1|^2\mb P_\perp \right\}_{\mb E_{\text{aniso}}^{(0)}},
\end{equation}
where the braced notation delineates the explicit assignment of terms to their respective basis spaces.

We next evaluate the elastic tensor Laplacian \(\nabla^2\mb Q_0\). Since the baseline director conforms to the constraint \(\mb n_0\cdot\nabla\mb n_0=\mb 0\), the spatial gradient tensor \(\nabla\mb n_0\) acts entirely within the transverse tangent plane. Consequently, the final tensor product in~\eqref{lap1} can be partitioned into a traceless transverse piece and a scalar component matching the uniaxial axis:
\begin{equation}
    2(\nabla\mb n_0)(\nabla\mb n_0)^T = 2\left[(\nabla\mb n_0)(\nabla\mb n_0)^T-\tfrac{1}{2}|\nabla\mb n_0|^2\mb P_\perp\right] + \left(\tfrac{2}{3}\mb I-\mb E_\parallel\right)|\nabla \mb n_0|^2.
\end{equation}
Similarly, differentiating the constant-length constraint yields the geometric identity \(\mb n_0\cdot\nabla^2\mb n_0 = -|\nabla \mb n_0|^2\), which permits the decomposition of the vector Laplacian into parallel and normal projections: \(\nabla^2\mb n_0=-|\nabla\mb n_0|^2\mb n_0 + \mc P_\perp(\nabla^2\mb n_0)\). Substituting these identities back into the linear distortion terms leads to the expression
\begin{align}
    (\nabla^2\mb n_0)\otimes\mb n_0 + \mb n_0\otimes(\nabla^2\mb n_0)=-2|\nabla\mb n_0|^2\mb E_\parallel-\tfrac{2}{3}|\nabla \mb n_0|^2\mb I  + \mc P_\perp(\nabla^2\mb n_0)\otimes\mb n_0 + \mb n_0\otimes \mc P_\perp(\nabla^2\mb n_0).
\end{align}
Gathering these components under their proper structural annotations yields the complete basis projection for the elastic forcing tensor:
\begin{align}
    \nabla^2\mb Q_0 &= -3|\nabla\mb n_0|^2\mb E_\parallel + \left\{\mc P_\perp(\nabla^2\mb n_0)\otimes\mb n_0 + \mb n_0\otimes \mc P_\perp(\nabla^2\mb n_0)\right\}_{\mb E_{tilt}^{(2)}} \nonumber \\
    &\quad + \left\{2\left[(\nabla\mb n_0)(\nabla\mb n_0)^T-\tfrac{1}{2}|\nabla\mb n_0|^2\mb P_\perp\right] \right\}_{\mb E_{aniso}^{(0)}}.
\end{align}

Equating the individual sub-spaces across the system reveals the definitive structural responses of the medium. First, isolating the geometric tilt subspace fields provides the essential solvability condition at this order:
\begin{equation}
\mc P_\perp(\nabla^2\mb n_0)\otimes\mb n_0 + \mb n_0\otimes \mc P_\perp(\nabla^2\mb n_0)=\mb 0 \qquad \implies \qquad  \mc P_\perp(\nabla^2\mb n_0)=\mb 0.
\end{equation}
From this relation, we recover identically the classical Oseen--Frank non-linear equilibrium equation system for the baseline director axis,
\begin{align}
\nabla^2\mb n_0 + |\nabla\mb n_0|^2\mb n_0=\mb 0 \quad \text{in} \quad \Omega,\qquad\mb n_0 = \mb n_b \quad \text{on} \quad \partial\Omega,
\end{align}
which confirms that the leading-order field operates under conventional Dirichlet boundary conditions. Next, balancing the components acting within the uniaxial scalar subspace yields a deterministic algebraic formula for the secondary scalar correction parameter
$s_2$:
\begin{equation}
    s_2 =  -\tfrac{3}{2}|\mb p_1|^2 -3\left(\tfrac{2}{3}-\mc A\right)^{-1}|\nabla\mb n_0|^2.
\end{equation}
Lastly, balancing the components acting within the transverse biaxial subspace yields the tensor mapping relation
\begin{equation}
    r_2 (\mb m_0\otimes\mb m_0-\tfrac{1}{2}\mb P_\perp) = \left\{ \mb L \right\}_{\mb E_{\text{aniso}}^{(0)}},
\end{equation}
wherein the symmetric forcing tensor $\mb L$ is explicitly defined by
\begin{equation}
    \mb L = \left(\mb p_1\otimes\mb p_1-\tfrac{1}{2}|\mb p_1|^2\mb P_\perp\right) + \frac{2}{\mc B}\left[(\nabla\mb n_0)(\nabla\mb n_0)^T-\tfrac{1}{2}|\nabla\mb n_0|^2\mb P_\perp\right].
\end{equation}
This relation delivers a dual set of structural parameters, simultaneously determining the secondary anisotropy amplitude $r_2$ and the baseline transverse orientation vector $\mb m_0$. Given that $\mb L$ is inherently traceless and resides completely within the local tangent manifold, satisfying the property $\mb P_\perp \mb L\mb P_\perp = \mb L$, it admits a  representation of the form $\mb L = \rho (\gb\mu\otimes\gb\mu-\tfrac{1}{2}\mb P_\perp)$. Here, $\rho = \sqrt{2\tr\mb L^2}$ denotes the positive eigenvalue of the configuration tensor, and $\gb\mu$ represents its corresponding principal eigenvector satisfying $|\gb\mu|^2=1$. By direct matching, we establish the simple algebraic links
\begin{equation}
    r_2 = \rho, \qquad \mb m_0 = \gb\mu.
\end{equation}

Thus far, we have derived the closed system of governing equations for the leading director field $\mb n_0$ and expressed the secondary perturbations to the core uniaxial state $(s=1, r=0)$ as functionals of the lower-order fields $\mb p_1$ and $\mb n_0$. The singular remaining unknown in this multi-scale formulation is the vector tilt perturbation $\mb p_1$, which must be resolved by ascending to the next stage of the asymptotic hierarchy.

\subsection{Third order}\label{sec:thirdorder}

At the tertiary asymptotic stage, the governing differential relations simplify to the non-homogeneous system
\begin{equation}\label{thirdordereq}
   \gb\Lambda_{\mb Q_0}'(\mb Q_3) = - \gb\Lambda_{\mb Q_0}''(\mb Q_1,\mb Q_2)  - \tfrac{1}{6} \gb\Lambda_{\mb Q_0}'''(\mb Q_1,\mb Q_1,\mb Q_1) + \nabla^2\mb Q_1.
\end{equation}
To evaluate the existence criteria for the third-order configuration tensor $\mb Q_3$, we project the entire system onto the structural tilt subspace $\mb E_{\text{tilt}}^{(3)}$. This operation isolates the underlying compatibility framework by completely annihilating the linear operator term on the left-hand side, thereby yielding a definitive solvability condition for the expansion.

By exploiting the geometric constraints $\mb p_1\cdot\mb n_0=0$ and the non-linear director Laplacian identity $\nabla^2\mb n_0=-|\nabla\mb n_0|^2\mb n_0$, a component-by-component projection shows that the linear elastic diffusion vector maps onto the tilt manifold according to
\begin{equation}
    \left[\nabla^2\mb Q_1\right]_{\text{tilt}} = \mc P_\perp(\nabla^2\mb p_1)-|\nabla\mb n_0|^2\mb p_1 -2 \left[(\nabla\mb n_0)(\nabla\mb n_0)^T\cdot\mb p_1\right].
\end{equation}
Furthermore, the corresponding projections of the higher-order cubic potential derivatives yield the following structural interactions:
\begin{align}
    &\left[- \tfrac{1}{6} \gb\Lambda_{\mb Q_0}'''(\mb Q_1,\mb Q_1,\mb Q_1)\right]_{\text{tilt}} = -2|\mb p_1|^2\mb p_1,\\
    &\left[- \gb\Lambda_{\mb Q_0}''(\mb Q_1,\mb Q_2)\right]_{\text{tilt}} = \frac{\mc B-4}{3}s_2\mb p_1  + \mc Br_2(\mb m_0\otimes\mb m_0-\tfrac{1}{2}\mb P_\perp)\mb p_1 .
\end{align}
Upon substituting these explicit representations into the global projection, and invoking the analytical forms previously derived for the scalar modification $s_2$ and the transverse anisotropy parameter $r_2$, a satisfactory cancellation occurs. The intricate non-linear couplings and material parameters cross-cancel each other identically throughout the bulk, collapsing the system down to an elegant linear boundary value problem:
\begin{align}
    \mc P_\perp(\nabla^2\mb p_1) + |\nabla\mb n_0|^2\mb p_1 = \mb 0 \quad \text{in} \quad \Omega, \qquad 
    \mb p_1 = -\frac{1}{\gamma}\frac{\partial\mb n_0}{\partial\nu} \quad \text{on} \quad \partial\Omega.
\end{align}
Here, the boundary condition emerges naturally from the first-order expansion of the surface anchoring relation.

We have thus established that the leading-order tilt correction obeys the classical homogeneous Jacobi equation of the limiting harmonic map, restricted strictly to the local tangent bundle and entirely decoupled from the non-linear biaxial fields that appear at intermediate stages of the expansion. This result is both physically satisfying and mathematically harmonious; since the tilt field $\mb p_1$ is constructed as an infinitesimal perturbation of the principal director axis $\mb n_0$, it is natural that it should conform to the linearised (Jacobi) version of the field equation which governs $\mb n_0$ itself, completely immune to the structural variations of the core bulk nematic potentials.

\section{Summary of the results}\label{sec:summary}

We now proceed to summarise the structural results emerging from our  asymptotic analysis. In the limit $\ep = \xi / h \to 0$, the spatial configuration of the tensor order parameter $\mb Q$ outside core neighbourhoods yields the formal representation
\begin{align}
    \mb Q &= (1+\ep^2s_2)\left(\mb n_0\otimes\mb n_0-\tfrac{1}{3}\mb I\right) + \ep\left(\mb p_1\otimes \mb n_0+\mb n_0\otimes\mb p_1\right) + \ep^2\left(\mb p_2\otimes \mb n_0+\mb n_0\otimes\mb p_2\right) \nonumber \\ 
    &\quad + \ep^2 r_2\left(\mb m_0\otimes\mb m_0-\tfrac{1}{2}\mb P_\perp\right) + \mc O(\ep^3),
\end{align}
wherein the local projection tensor onto the transverse plane is denoted by $\mb P_\perp = \mb I - \mb n_0 \otimes \mb n_0$. The baseline Oseen--Frank director field $\mb n_0(\mb x)$, subject to the constraint $|\mb n_0|^2=1$, satisfies the nonlinear Dirichlet problem
\begin{align} \label{directorproblem}
    \nabla^2\mb n_0 + |\nabla\mb n_0|^2\mb n_0 = \mb 0 \quad \text{in} \quad \Omega,\qquad
    \mb n_0 = \mb n_b \quad \text{on} \quad \partial\Omega.
\end{align}
The leading-order vector tilt perturbation $\mb p_1(\mb x)$ is governed by the system
\begin{align} \label{tiltproblem}
    -\mc J_{\mb n_0}(\mb p_1) \equiv \mc P_\perp(\nabla^2\mb p_1) + |\nabla\mb n_0|^2\mb p_1 = \mb 0 \quad \text{in} \quad \Omega,\qquad 
    \mb p_1 = -\frac{1}{\gamma}\frac{\partial\mb n_0}{\partial\nu} \quad \text{on} \quad \partial\Omega,
\end{align}
where $\mc J_{\mb n_0}$ is the on-shell $(\mathbb S^2)$ linear Jacobi operator of the limiting harmonic map $\mb n_0$. We note that an alternative choice of surface anchoring mechanics would modify exclusively the functional structure of the right-hand side of the boundary condition in~\eqref{tiltproblem}. The secondary correction $s_2$ modifying the bulk uniaxial scalar state evaluates to the explicit form
\begin{equation}
    s_2 =  -\tfrac{3}{2}|\mb p_1|^2 -3\left(\tfrac{2}{3}-\mc A\right)^{-1}|\nabla\mb n_0|^2.
\end{equation}
The matching transverse biaxial tensor component $r_2\left(\mb m_0\otimes\mb m_0-\tfrac{1}{2}\mb P_\perp\right)$, aligned with the secondary axis field $\mb m_0=(\cos\psi,\sin\psi,0)^T$, is completely determined by the relations
\begin{align}
    r_2 = \sqrt{2\tr\mb L^2}, \qquad \psi = \frac{1}{2}\mathrm{atan2}\left(L_{12},{L_{11}}\right),
\end{align}
wherein the symmetric elastic interaction tensor $\mb L$ is defined as
\begin{equation} \label{Lequation}
    \mb L = \left(\mb p_1\otimes\mb p_1-\tfrac{1}{2}|\mb p_1|^2\mb P_\perp\right) + \frac{2}{\mc B}\left[(\nabla\mb n_0)(\nabla\mb n_0)^T-\tfrac{1}{2}|\nabla\mb n_0|^2\mb P_\perp\right].
\end{equation}

Manifestly, there exists a secondary vector correction to the tilt field, denoted by $\mb p_2$, within the general expansion hierarchy of the order parameter tensor, which remains undetermined within the current scope. By extension, this next-order field is expected to satisfy a non-homogeneous system of the form $\mc P_\perp(\nabla^2\mb p_2) + |\nabla\mb n_0|^2\mb p_2 = \mc F(s_2, r_2)$, where $\mc F$ represents a prescriptive forcing term. This structural framework is consistent with the behavior recovered in reduced two-dimensional formulations~\cite{rajamanickam2026strong}. We do not pursue this rather cumbersome calculation here owing to a fundamental invariance requirement: a geometric director tilt of order $\mc O(\ep^k)$ must invariably be accompanied by matching shifts in the parameters $s$ and $r$ at order $\mc O(\ep^{k+1})$ to preserve the rotational invariance of the baseline uniaxial configuration in spatially uniform regimes ($\nabla\mb n_0 = \mb 0$). The transverse anisotropy tensor $\mb L$ and the bulk scalar correction $s_2$ both receive separate contributions on an equal footing, driven concurrently by the non-linear self-coupling of the tilt field $\mb p_1$ and the elastic distortions $\nabla\mb n_0$.

To delineate the spectral variations within the medium, it is instructive to evaluate the eigenvalues of the global $\mb Q$-tensor. Without loss of generality, adopt a localised Cartesian reference frame aligned such that $\mb n_0 = \mb e_z$ and $\mb m_0 = \mb e_x$. In this coordinate  system, the components of the $\mb Q$-tensor rearrange into the matrix form
\begin{align}
    \mb Q = \begin{pmatrix}
        -\tfrac{1}{3}(1+\ep^2s_2)+\tfrac12 \ep^2 r_2 & 0 & \ep p_x \\ 0 & -\tfrac{1}{3}(1+\ep^2 s_2)-\tfrac12 \ep^2 r_2 & \ep p_y \\ \ep p_x & \ep p_y & \tfrac{2}{3}(1+\ep^2s_2)
    \end{pmatrix}.
\end{align}
Solving the corresponding characteristic equation $\det(\mb Q - \lambda \mb I) = 0$ yields the three coordinate-free eigenvalues up to order $\mc O(\ep^2)$:
\begin{align}
    \lambda_1 &= \frac{2}{3} -2\ep^2 \left(\tfrac{2}{3}-\mc A\right)^{-1}|\nabla\mb n_0|^2, \\
    \lambda_{2,3} &= -\frac{1}{3}+\ep^2\left\{\left(\tfrac{2}{3}-\mc A\right)^{-1}|\nabla\mb n_0|^2 \pm \frac{\sqrt2}{B} \left|(\nabla\mb n_0)(\nabla\mb n_0)^T-\tfrac12|\nabla\mathbf n_0|^2\mathbf P_\perp\right|\right\}.
\end{align}
Remarkably, the tilt field $\mb p_1$ drops out of all three eigenvalues at this order. Its contributions to $s_2$ and to $\mb L$ combine into the second-order expansion of a rigid rotation of the uniaxial state, which leaves the spectrum unchanged. The eigenvalues are therefore determined solely by the elastic distortion $\nabla\mb n_0$: the principal eigenvalue $\lambda_1$ and the mean of $\lambda_{2,3}$ are set by $|\nabla\mb n_0|^2$, while the splitting $|\lambda_2-\lambda_3|=\mc O(\ep^2)$ is controlled by the traceless part of $(\nabla\mb n_0)(\nabla\mb n_0)^T$. Consequently, the biaxiality parameter $\beta = 1 - 6(\tr \mb Q^3)^2/(\tr \mb Q^2)^3 \sim (\lambda_2-\lambda_3)^2$ is $\mc O(\ep^4)$, which is a minor correction, and it is generated purely by the anisotropy of the elastic distortion. In particular, $\beta$ vanishes identically when $\nabla\mb n_0=\mb 0$, as it must, since a uniform tilt is a rigid rotation of the uniaxial state.

Integrating the total free energy functional~\eqref{freeenergy} over the domain yields the global energetic response, which expands apart from an additive constant as
\begin{align}
    F = \int_\Omega |\nabla\mb n_0|^2 dV - \ep\gamma\int_{\partial\Omega}\vert \mb p_1\vert^2d\Sigma + \ep^2\left(\int_{\Omega}f_2dV + \int_{\partial\Omega}\mb p_1\cdot\pfr{\mb p_1}{\nu} d\Sigma \right) + \cdots,
\end{align}
wherein the simplified second-order energy density is given by
\begin{align}
    f_2 =  -\frac{2}{\mc B}\left\vert (\nabla\mb n_0)(\nabla\mb n_0)^T-\tfrac{1}{2}|\nabla\mb n_0|^2\mb P_\perp\right\vert^2 - \frac{9}{4-\mc B}|\nabla \mb n_0|^4 .
\end{align}
First of all, it is a satisfactory result that the tilt field does not enter the bulk energy since in homogeneous systems where $\nabla\mb n_0=\mb 0$, the tilt being a rigid-body rotation should not cost any energy. Furthermore, since the linear coefficient governing the first-order correction at the boundary is strictly negative, the physical liberation of these soft tilt modes provides a direct energetic pathway for the system to shed elastic energy at its bounding walls. The sign of the second-order surface term is given by $\int_{\partial\Omega}d\Sigma \mb p_1\cdot\pfi{\mb p_1}{\nu}=\int_{\Omega}dV[|\nabla\mb p_1|^2-|\nabla \mb n_0|^2|\mb p_1|^2]\geq 0$ whenever $\sigma_1>0$ (see Section~\ref{sec:solvability} below).

\section{Discussions}\label{sec:discussion}

\subsection{An illustrative example: Corrections to a pure twist}\label{sec:twist}

To illustrate the concrete physical implications of the foregoing  asymptotic theory, we consider a nematic liquid crystal sample subjected to a pure torsional deformation, characterized by a uniform twist angle $\alpha$ per unit axial length. This configuration is sustained by two parallel bounding planes situated at $z=0$ and $z=1$. Adopting a cylindrical coordinate system $(r,\theta,z)$ with the azimuthal orientation defined by $\theta=\alpha z$, the baseline Oseen--Frank director distribution is given by the radial vector field~\cite{de1995physics}
\begin{equation}
   \mb n_0 = \mb e_r, \qquad \mb e_r = \begin{pmatrix}
        \cos(\alpha z) \\ \sin(\alpha z) \\ 0
    \end{pmatrix}. 
\end{equation}
The general representation for the first-order tilt perturbation may be formulated as the linear combination
\begin{equation}
   \mb p_1 = u(z) \mb e_\theta + v(z) \mb e_z, \qquad  \mb e_\theta = \begin{pmatrix}
        -\sin(\alpha z) \\ \cos(\alpha z) \\ 0
    \end{pmatrix},
\end{equation}
wherein $u(z)$ and $v(z)$ denote the localised in-plane and out-of-plane tilt components, respectively. By substitution into the homogeneous bulk operator equations~\eqref{tiltproblem}, the  tilt components are found to satisfy the system of ordinary differential equations
\begin{align}
    \frac{d^2u}{dz^2} = 0,\qquad \frac{d^2v}{dz^2}  +\alpha^2 v = 0.
\end{align}
This system is subject to the boundary requirements $u(0)=\alpha/\gamma$, $u(1)=-\alpha/\gamma$, and $v(0)=v(1)=0$, derived directly from the linear expansion of the surface anchoring conditions. It follows immediately that the out-of-plane amplitude vanishes identically throughout the domain ($v(z) \equiv 0$). The unique configuration for the tilt field is thus delivered by the linear profile
\begin{equation}
    \mb p_1 = \frac{\alpha}{\gamma}(1-2z) \mb e_\theta.
\end{equation}
The tilt vector is seen to vanish precisely at the mid-plane of the domain ($z=1/2$) and increases monotonically toward the boundaries, peaking symmetrically at the walls with equal and opposite spatial orientations. The net physical consequence of this anchoring-driven relaxation is a global reduction in the net torsional twist of the macrostructure. The effective twist wavenumber within the medium evaluates to
\begin{equation}
    \alpha_{\mathrm{eff}} = \frac{d}{dz}\left[\alpha + \frac{\ep\alpha}{\gamma}(1-2z)\right] = \alpha \left(1-\frac{2\ep}{\gamma}\right).
\end{equation}
The  freedom at the boundary interface thus operates to globally soften the effective elastic response of the medium, exactly as if the physical cell were subjected to a lower nominal twist threshold, $\alpha_{\mathrm{eff}} < \alpha$, throughout its bulk. Furthermore, the geometric axes of the system align perfectly, satisfying the parallel condition $\mb p_1 \parallel \mb m_0 $, whereupon the secondary corrections map onto the explicit coordinate expressions
\begin{align}
    s_2 = -3\alpha^2\left[\frac{(1-2z)^2}{2\gamma^2}+\frac{1}{\tfrac{2}{3}-\mc A}\right], \qquad  r_2 = \alpha^2\left[\frac{(1-2z)^2}{\gamma^2}+\frac{2}{\mc B}\right].
\end{align}

In the construction detailed above, we have implicitly assumed that the total twist angle $\alpha$ does not coincide with an integer multiple of $\pi$, thereby ensuring the unique trivial solution for the out-of-plane component $v(z)$. However, when the system reaches the critical threshold values $\alpha = k\pi$ for $k=1,2,3,\dots$, the boundary value problem undergoes a fundamental structural shift. At these critical junctions, the out-of-plane field admits the non-trivial harmonic solution $v(z) = c_k \sin(k\pi z)$, yielding a generalized tilt field of the form
\begin{equation}
    \mb p_1 = \frac{\alpha}{\gamma}(1-2z) \mb e_\theta +  c_k \sin(k\pi z)\mb e_z,
\end{equation}
which is no longer unique owing to the arbitrary nature of the amplitude constants $c_k$. Mathematically, the limiting harmonic maps $\mb n_0$ corresponding to these discrete twist rates constitute the conjugate points of the variational system. Given that the appearance of such conjugate configurations marks the threshold of linear energetic instability, thus $\alpha=\pi$ marks the onset of the instability, beyond which the twisted state $\mb n_0=\mb e_r$ is no longer a local minimizer.

\subsection{Solvability condition for $\mb p_1$}\label{sec:solvability}

The illustrative example detailed above demonstrates concretely how the uniqueness of the vector tilt perturbation $\mb p_1$ can fail; we now proceed to formalise this structural compatibility criterion generally. The non-linear harmonic map problem itself frequently admits a multiplicity of solution branches on account of the head-to-tail inversion symmetry of the boundary data $\mb n_b \sim -\mb n_b$. Consequently, a fundamental question arises: once a specific baseline configuration $\mb n_0$ has been selected, does a physically valid solution for the matching tilt field $\mb p_1$ exist? To resolve this query, we must examine the spectral characteristics and mapping properties of the on-shell Jacobi operator $\mc J_{\mb n_0}$.

The determination of solvability via the classical Fredholm alternative requires a rigorous examination of the null-space kernel of the operator under homogeneous Dirichlet boundary constraints, defined as
\begin{equation}
    \mc K = \left\{\mb v \in H_0^1(\Omega,\mathbb R^3) \;\middle\vert{}\; \mb v\cdot\mb n_0 = 0 \quad \text{and} \quad \mc J_{\mb n_0}(\mb v)=\mb 0 \right\}.
\end{equation}
By evaluating the weak formulation of the constrained system, and applying Green's second identity to the components of $\mb p_1$ and an arbitrary kernel vector $\mb v \in \mc K$, the requirement that the test function vanishes at the boundary, $\mb v |_{\partial\Omega}=\mb 0$, yields the  global compatibility relation,
\begin{equation}
    \int_{\partial\Omega}\left(\mb v\cdot\frac{\partial\mb p_1}{\partial\nu}-\mb p_1\cdot\frac{\partial\mb v}{\partial\nu}\right)d\Sigma = 0 \quad \implies \quad \int_{\partial\Omega} \mb p_1\cdot\frac{\partial\mb v}{\partial\nu} d\Sigma = 0.
\end{equation}

In the non-degenerate regime where the kernel is empty, $\mc K = \{\mb 0\}$, this global solvability criterion is trivially  satisfied. This guarantees the existence of a unique, well-behaved solution for the tilt field $\mb p_1$ under any arbitrary surface anchoring distribution driven by the normal derivative of the base director field. Conversely, in the degenerate regime where $\mc K \neq \{\mb 0\}$, the system possesses non-trivial zero-energy  modes, that can be activated at zero bulk elastic energy cost. In this case, the linear system admits either infinitely many solutions or no solution at all for $\mb p_1$, depending strictly on whether the surface integrals satisfy the boundary compatibility condition. When this condition is met, the general field representation takes the form
\begin{equation}
    \mb p_1 = \mb p_{\text{part}} + \sum_k c_k \mb v_k,
\end{equation}
wherein $\mb p_{\text{part}}$ represents the particular solution satisfying the non-homogeneous anchoring boundary condition, and $\{\mb v_k\} \in \mc K$ denote the linearly independent zero-energy modes. This corresponds precisely to the non-unique state spaces mapping onto the discrete conjugate boundaries explored in the pure twist example. In more general, asymmetric three-dimensional scenarios  with $\mc K\neq\{0\}$, however, the normal stress variations at the wall will generically fail to satisfy the solvability condition. Once the compatibility integral is violated, a continuous solution for $\mb p_1$ ceases to exist. Physically, this failure signals a breakdown of the asymptotic $\mb Q$-tensor expansion, indicating that the baseline continuum field $\mb n_0$ has reached a bifurcation threshold.

\subsubsection{The eigenvalue problem}
From both an analytical and a numerical standpoint, tracking whether the null-space kernel is non-empty is best framed as a constrained eigenvalue problem:
\begin{equation}\label{eigenvalue}
     \mc J_{\mb n_0}(\mb v) = \sigma \mb v \quad \text{in} \quad \Omega,
\end{equation}
subject strictly to the geometric constraint $\mb v\cdot \mb n_0 = 0$ and the homogeneous boundary requirement $\mb v = \mb 0$ on $\partial\Omega$. The kernel $\mc K$ is non-empty if and only if zero is an element of the discrete spectrum. Assuming that the eigenvalues are ordered as in $\sigma_1 \leq \sigma_2 \leq \sigma_3 \leq \cdots$, we can classify the structural mechanics of the medium into three distinct regimes:
\begin{itemize}
    \item $\sigma_1 > 0$: The global spectrum is strictly positive definite. The lowest energy state under rigid boundaries requires the vanishing configuration $\mb v = \mb 0$, from which it follows that $\mc K = \{\mb 0\}$. A unique and stable tilt correction field $\mb p_1$ is mathematically guaranteed.
    \item \mbox{$\sigma_1 = 0$}: The system has reached a critical structural transition threshold (conjugate point). It costs zero bulk elastic energy to perturb the director field along the path of the fundamental eigenmode $\mb v_1$. The kernel becomes non-empty, and the explicit boundary compatibility integral must be evaluated to determine existence.
    \item $\sigma_1 < 0$: The appearance of negative eigenvalues indicates that the selected baseline state $\mb n_0$ constitutes an energy saddle point or a local maximum. The system is physically unstable under perturbations in the bulk.
\end{itemize}
It is critical to observe that the condition $\sigma_1 = 0$ marks the threshold of stability, not the onset of instability; the field $\mb n_0$ may remain a valid, albeit degenerate, local energy minimizer at this point. Whether the perturbative scheme survives this threshold is a separate structural question concerning the very existence of the tilt field $\mb p_1$, a question that is settled exclusively by the boundary solvability condition.

 Writing $\mb D = P_\perp \nabla$ and
$\Delta^T = \mb D\cdot\mb D$ for the covariant Laplacian on the tangent bundle, one has
$P_\perp\nabla^2\mb v = \Delta^T\mb v - (\nabla\mb n_0)(\nabla\mb n_0)^T\mb v$,
so~\eqref{eigenvalue} becomes
\begin{equation}
    -\Delta^T\mb v + \mb V\mb v = \sigma\mb v,\qquad
    \mb V = -\left(|\nabla\mb n_0|^2 \mb I - (\nabla\mb n_0)(\nabla\mb n_0)^T\right).
\end{equation}
This is formally a stationary Schr\"odinger operator on the tangent bundle, with a connection
arising from the projection and a matrix-valued potential coming from the curvature of
$\mathbb S^2$. The potential is negative semidefinite: for any unit vector $\mb w$,
$\mb w\cdot(\nabla\mb n_0)(\nabla\mb n_0)^T\mb w=\sum_j(\mb w\cdot\partial_j\mb n_0)^2\le|\nabla\mb n_0|^2$
by the Cauchy--Schwarz inequality. Equivalently, since $\mb n_0\cdot\partial_j\mb n_0=0$, the
matrix $(\nabla\mb n_0)(\nabla\mb n_0)^T$ annihilates $\mb n_0$ and maps the tangent plane into
itself, where it is positive semidefinite with eigenvalues $g_1,g_2\ge0$ satisfying
$g_1+g_2=|\nabla\mb n_0|^2$. On the tangent plane, the eigenvalues of $\mb V$ are therefore
$-g_2$ and $-g_1$, both non-positive. Thus $\mb V$ is an attractive well, deepest where the
director bends most, although its depth can differ between the two tangent directions.
The Dirichlet Laplacian alone is positive, so $\sigma_1>0$ when the well is shallow relative
to the confinement, while a sufficiently deep or localised well produces bound states with
$\sigma_1<0$. The regimes above are thus the usual competition between confinement and well
depth.

\section{Planar configurations and escape to the third dimension}\label{sec:numerical}

Suppose the baseline director is confined to the $xy$-plane, $\mb n_0=(\cos\varphi,\sin\varphi,0)^T$,
and let $\mb e_\theta=(-\sin\varphi,\cos\varphi,0)^T$. Under this planar constraint, the nonlinear
harmonic map equation~\eqref{directorproblem} reduces to Laplace's equation for the orientation angle,
\begin{equation}
    \nabla^2\varphi=0 \quad \text{in } \Omega,\qquad \varphi=\varphi_b \quad \text{on } \partial\Omega .
    \label{planarphi}
\end{equation}
Since $P_\perp\partial_j\mb e_\theta=\mb 0$ and $\mb e_z$ is constant, both tangent vectors are
covariantly constant, so that
$\Delta^T(u\,\mb e_\theta+v\,\mb e_z)=(\nabla^2u)\,\mb e_\theta+(\nabla^2v)\,\mb e_z$.
Moreover $(\nabla\mb n_0)(\nabla\mb n_0)^T=|\nabla\varphi|^2\,\mb e_\theta\otimes\mb e_\theta$, so that
$\mb V=-|\nabla\varphi|^2\,\mb e_z\otimes\mb e_z$. Writing the tilt as
\begin{equation}
    \mb p_1=u(x,y)\,\mb e_\theta+v(x,y)\,\mb e_z ,
\end{equation}
the Jacobi problem~\eqref{tiltproblem} therefore decouples into
\begin{equation}
    \nabla^2u=0,\quad \nabla^2 v+|\nabla\varphi|^2 v=0 \quad \text{in } \Omega,\qquad
    u=-\frac{1}{\gamma}\frac{\partial\varphi}{\partial\nu},\quad v=0 \quad \text{on } \partial\Omega .
    \label{planardecoupled}
\end{equation}
The in-plane component $u$ sees no potential and obeys a free Dirichlet Laplacian, which is strictly
positive definite; it is therefore uniquely solvable for any smooth boundary forcing. The out-of-plane
component $v$ sees the attractive well $-|\nabla\varphi|^2$. Consequently, a planar harmonic map can
lose stability, or admit a non-trivial kernel, only through the $\mb e_z$ channel, that is, by escaping
into the third dimension.  To delineate solvability and
stability we thus need only the scalar eigenvalue problem
\begin{equation}
    \nabla^2 v+|\nabla\varphi|^2 v = -\sigma v \quad \text{in } \Omega,\qquad
    v=0 \quad \text{on } \partial\Omega .
    \label{planareig}
\end{equation}
As a corollary, in the reduced Landau--de Gennes framework, where $\mb n_0\in\mathbb S^1$ and the
$\mb e_z$ channel is absent by construction, the Jacobi operator reduces to the Dirichlet Laplacian.
Hence $\sigma_1>0$ for every harmonic map, the kernel $\mathcal K$ is trivial, the tilt field $\mb p_1$
exists and is unique for every $\gamma$, and $\mb n_0$ is stable at this order.

\begin{figure}[h!]
\centering
\includegraphics[width=0.7\textwidth]{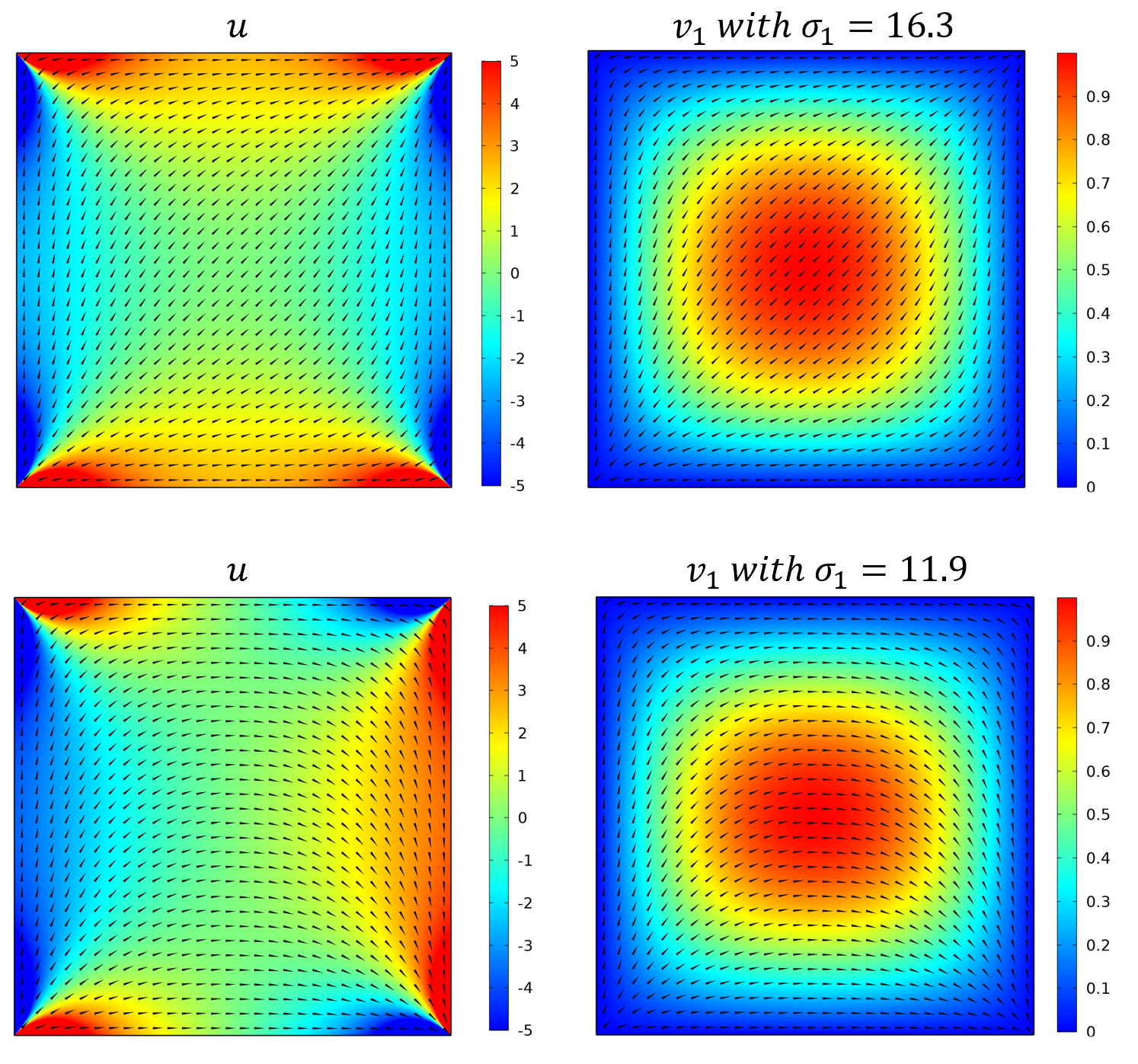}
\caption{Numerical solutions for the stable structural states: the top row delineates the canonical diagonal (D) configuration, and the bottom row illustrates the rotated (R) configuration. The baseline director orientation fields are established under prescribed boundary data: $\varphi_b=0$ at the horizontal walls and $\varphi_b=\pi/2$ at the vertical walls for the D-state; and $\varphi_b=0$ at the horizontal walls, matched with $\varphi_b=\pi/2$ (left) and $\varphi_b=-\pi/2$ (right) for the R-state. The corresponding in-plane tilt distributions $u(x,y)$ are detailed in the left panels, with contours restricted to the interval $[-5,5]$ to isolate the corner singularities. The primary stabilising eigenfunctions $v_1(x,y)$ are detailed in the right panels.} 
\label{fig:stable}
\end{figure}

\begin{figure}[h!]
\centering
\includegraphics[width=0.7\textwidth]{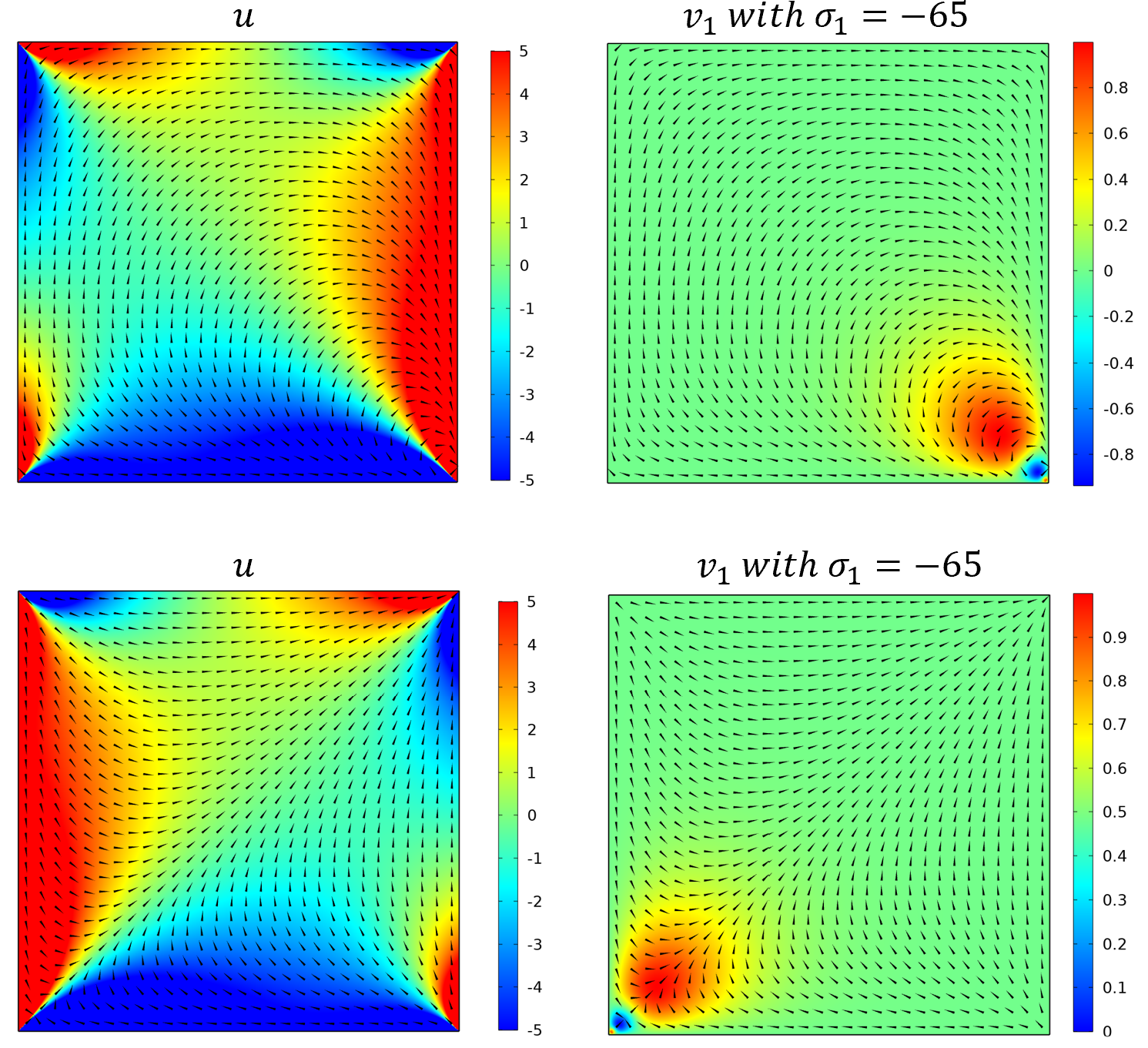}
\caption{Numerical solutions for the unstable critical branches: the plots delineate a contrasting pair of non-minimizing tangential configurations characterized by a negative eigenvalue ($\sigma_1 < 0$). The upper row illustrates the solution branch generated under boundary parameters $\varphi_b=\pi$ (bottom), $\varphi_b=0$ (top), $\varphi_b=\pi/2$ (left), and \mbox{$\varphi_b=-\pi/2$} (right). The lower row delineates the corresponding state under configurations $\varphi_b=\pi$ (bottom), $\varphi_b=0$ (top), $\varphi_b=-\pi/2$ (left), and $\varphi_b=\pi/2$ (right). The localised pattern seen in the unstable eigenfunction near the corner, which is more singular than the usual bend or splay corner~\cite{rajamanickam2026nematic,han2020reduced}, indicates an escape to the third dimension. The value $\sigma_1=-65$ is specific to the mesh used: because the phase jump at the corner exceeds $\pi$, the Jacobi form is unbounded below near the corner and $\sigma_1$ does change with the mesh size.} 
\label{fig:unstable}
\end{figure}

We now illustrate these results numerically in a square cylinder of unit side length with tangential
anchoring, where $\varphi_b$ is piecewise constant, prescribed edge by edge. The boundary value
problem~\eqref{planarphi} admits an analytical representation via separation of variables~\cite{majumdar2016multistable},
but we proceed here by direct numerical evaluation of~\eqref{planarphi}--\eqref{planareig}.

We consider first the canonical topologically distinct configurations, namely, the diagonal (D) state and the rotated (R) state. Both configurations constitute well-known, energetically stable local minimizers of the full Landau--de Gennes free energy functional within the asymptotic large-domain limit ($\ep \to 0$). The computed baseline director fields, alongside their unique in-plane tilt profiles $u$ and fundamental eigenfunctions $v_1$, are detailed in Figure~\ref{fig:stable}. For both baseline branches, the discrete spectrum of eigenvalues $\{\sigma_n\}$ is found to remain strictly positive definite, yielding the fundamental ground-state metrics $\sigma_1 = 16.3$ for the diagonal state and \mbox{$\sigma_1 = 11.9$} for the rotated state. This positive-definite character establishes that the null-space kernel is empty, $\mc K = \{\mb 0\}$, across both configurations. Consequently, a unique and stable vector tilt correction field $\mb p_1$ is mathematically guaranteed to exist for any arbitrary surface anchoring metric $\gamma$. Furthermore, because \mbox{$\sigma_1 > 0$}, both configurations represent authentic local energy minimizers of the Oseen--Frank Dirichlet functional. The in-plane component $u$ (left panels in Fig.~\ref{fig:stable}) is finite everywhere except at the four corners, where the discontinuity in $\varphi_b$ between adjacent edges forces an inverse distance singularity, as already noted in~\cite{rajamanickam2026strong}.

A contrasting regime is illustrated in Figure~\ref{fig:unstable}, which details a secondary pair of solution branches derived from boundary alignments that remain strictly tangential but do not correspond to the  energy-minimizing states. These harmonic maps are still bona fide critical points of the Dirichlet energy, but the spectrum is now found to contain a negative leading eigenvalue. Thus, no eigenvalue in the computed spectrum passes through zero on the discretised problem and the kernel remains trivial, $\mc K=\{0\}$. Consequently, the tilt problem is uniquely solvable at every $\gamma$.
 However, $\sigma_1<0$ signals that $\mb n_0$ is an unstable point of the Dirichlet energy. Infinitesimal transverse perturbations directed along the principal eigenmode \(v_1\)---which is highly localised near the geometric corners in Figure~\ref{fig:unstable} owing to the massive local gradients \(|\nabla\varphi|^2\) that deepen the effective potential well---will operate to strictly lower the total bulk energy.  Nevertheless, this instability is entirely a statement about $\mb n_0$ itself, and is logically independent of the solvability condition for $\mb p_1$. The tilt correction $\mb p_1$ is therefore well defined and unique for this branch even though the underlying harmonic map is not a minimizer, but a saddle point or maximizer.  

\section{Concluding remarks}\label{sec:concluding}

In this work, we have presented a three-dimensional asymptotic framework to analyze higher-order Landau--de Gennes corrections to the Oseen--Frank limit under a physically appropriate scaling of the surface anchoring energy. By relaxing the infinite anchoring constraint implicit in traditional treatments, this formulation captures a rotationally invariant expansion that natively preserves the coupling between boundary variations and bulk elastic profiles. A central feature of this asymptotic expansion is the emergence of a non-vanishing tilt field $\mb p_1$ at the leading $\mc O(\ep)$ order. Under the conventional benchmark of rigid Dirichlet boundary conditions, the director is fixed at the walls across all orders, which effectively restricts these boundary responses from developing at $\mc O(\ep)$ and shifts the corrections into \mbox{$\mc O(\ep^2)$} bulk eigenvalue variations. By contrast, retaining a compliant surface energy at its physical scaling allows the director profile to respond to local elastic fields at the boundary interface. This relaxation acts as a direct source term that activates macroscopic, geometric tilt modes.

The algebraic structure of the third-order expansion reveals a clean isolation of this tilt field. Once the lower-order solutions for the uniaxial scalar modification $s_2$ and the transverse anisotropy parameter $r_2$ are introduced into the projection equations, the non-linear coupling contributions cancel out perfectly in the bulk. Consequently, the  tilt perturbation $\mb p_1$ is found to satisfy the homogeneous on-shell Jacobi equation of the limiting harmonic map, restricted entirely to the local tangent bundle. It is worth noting that such Jacobi equations have been studied in the past~\cite{virga2018variational}, but purely from the viewpoint of the Oseen--Frank theory, which resides strictly on the unit sphere $\mathbb S^2$. Our analysis shows that the tilt decouples from the nonlinear bulk fields in the full Landau--de Gennes framework, so it tracks the linearised continuum equation for $\mb n_0$.

The underlying structure of this correction follows directly from the rotational invariance of the bulk Landau--de Gennes energy functional. A rigid rotation of a uniaxial state leaves the bulk potential unchanged; hence, an infinitesimal reorientation of $\mb n_0$ incurs zero bulk potential energy. This degeneracy explains why the differential operator governing the tilt equation reduces strictly to the classical Jacobi equation of the harmonic map $\mb n_0$. In a confined sample, this rotational invariance is not spontaneously broken, as the bounding walls break it explicitly by preferring the fixed direction $\mb n_b$. The tilt is therefore not a gapless Goldstone excitation of the confined system, but rather a soft mode of the rotationally invariant bulk energy~\cite{landau1981statistical,de1969long}. Its exclusive restoring forces are the elastic response of the harmonic map and the surface anchoring at the wall.

The same reasoning governs how the tilt enters the free energy and the local order. A tilt of order $\ep$ is accompanied by matching $\mc O(\ep^2)$ corrections to the scalar $s_2$ and the anisotropy $\mb L$, which together form the second-order expansion of a rigid rotation of the uniaxial state. Through order $\mc O(\ep^2)$ these corrections cancel the tilt's own bulk cost, so that $\mb p_1$ manifests solely within surface integrals. For the same reason, $\mb p_1$ drops out of the eigenvalues of $\mb Q$ at this order, so the tilt generates no biaxiality through $\mc O(\ep^2)$; the biaxiality is produced entirely by the elastic anisotropy of $\nabla\mb n_0$. This holds to the order computed, and we make no claim beyond it. At order $\mc O(\ep)$, the wall term enters negatively as $-\ep\gamma\oint_{\partial\Omega}|\mb p_1|^2\,d\Sigma$, confirming that finite anchoring lowers the total energy by allowing the director field to relax from the baseline Oseen--Frank field at the wall.

Several open questions remain for future investigation. These include the generalisation to alternative anchoring laws beyond the classical Rapini--Papoular formulation (which would modify exclusively the boundary condition for $\mb p_1$), the systematic matching of the tilt fields to inner defect-core solutions, and the relaxation of the single-elastic-constant approximation. Finally, the explicit calculation of the second-order tilt vector $\mb p_2$, alongside its companion multi-scale scalar and anisotropy partners $(s_3, r_3)$, presents a natural pathway to extend the accuracy of the theory to higher asymptotic thresholds.





\bibliographystyle{ejamike}
\bibliography{refs}

\appendix

\section{Small domain limit, $\ep \to \infty$} \label{sec:small}

The analysis for the small-domain limit is essentially unchanged from that given in~\cite{rajamanickam2026strong} within the reduced Landau--de Gennes framework. For completeness, we provide the results for three-dimensions here. The solution is expanded by the series
\begin{equation}
    \mb Q = \sum_{m=0}^\infty \ep^{-m} \mb Q_m.
\end{equation}
At leading order, the solution is constant and is given by
\begin{align}
    \mb Q_0 = \frac{1}{|\partial\Omega|}\oint_{\partial\Omega}\mb Q_b\,d\Sigma .
\end{align}
At first order, we obtain
\begin{align}
    &\nabla^2 \mb Q_1=0, \quad \text{on } \Omega,\\
     &\pfr{\mb Q_1}{\nu} = -  \gamma(\mb Q_0 - \mb Q_b),  \quad \text{on} \quad \partial\Omega,\\
     &\gamma \oint_{\partial\Omega} \mb Q_1 d\Sigma = - \left[\mc A \mb Q_0 - \mc B (\mb Q_0^2 - \tfrac{1}{3}\mb I\tr\mb Q_0^2) + \mb Q_0\tr\mb Q_0^2\right] |\Omega|.
\end{align}
At higher orders ($m\geq 2$), we obtain 
\begin{align}
     &\nabla^2 \mb Q_m= \gb \Lambda_{m-2},  \quad \text{on } \Omega,\\
     &\pfr{\mb Q_m}{\nu} = -  \gamma \mb Q_{m-1},  \quad \text{on} \quad \partial\Omega,\\
     &\gamma\oint_{\partial\Omega} \mb Q_m d\Sigma = -\int_{\Omega} \gb\Lambda_{m-1} dV .
\end{align}
where $\gb\Lambda_m$ is the m-th order term of $ \mc A \mb Q - \mc B (\mb Q^2 - \tfrac{1}{3}\mb I\tr\mb Q^2) + \mb Q\tr\mb Q^2$.

\end{document}